\documentclass[graybox]{svmult}

\usepackage{mathptmx}       
\usepackage{helvet}         
\usepackage{courier}        
\usepackage{type1cm}        
\usepackage{makeidx}         
\usepackage{graphicx}        
\usepackage{multicol}        
\usepackage[bottom]{footmisc}
\usepackage{cite}
\usepackage{amsmath,amssymb,amsfonts}
\usepackage{graphicx}
\usepackage{textcomp}
\usepackage{xcolor}
\usepackage{graphicx}
\usepackage{subcaption}
\usepackage{csquotes}
\usepackage{multirow}
\usepackage[linesnumbered,ruled,vlined]{algorithm2e}

\SetCommentSty{mycommfont}

\SetKwInput{KwInput}{Input}                
\SetKwInput{KwOutput}{Output}              

\makeindex             

\begin{document}

\title*{On Network Design and Planning 2.0 for Optical-computing-enabled Networks}
\author{Dao Thanh Hai, Isaac Woungang}
\institute{Dao Thanh Hai \at School of Science, Engineering and Technology, RMIT University Vietnam \email{hai.dao5@rmit.edu.vn}
\and Isaac Woungang \at Department of Computer Science, Toronto Metropolitan University, Toronto, ON, Canada  \email{iwoungan@torontomu.ca}}
%
%
\maketitle

\abstract{In accommodating the continued explosive growth in Internet traffic, optical core networks have been evolving accordingly thanks to numerous technological and architectural innovations. From an architectural perspective, the adoption of optical-bypass networking in the last two decades has resulted in substantial cost savings, owning to the elimination of massive optical-electrical optical interfaces. In optical-bypass framework, the basic functions of optical nodes include adding (dropping) and cross-connecting transitional lightpaths. Moreover, in the process of cross-connecting transiting lightpaths through an intermediate node, these lightpaths must be separated from each other in either time, frequency or spatial domain, to avoid unwanted interference which deems to deteriorate the signal qualities. In light of recently enormous advances in photonic signal processing / computing technologies enabling the precisely controlled interference of optical channels for various computing functions, we propose a new architectural paradigm for future optical networks, namely, \textit{optical-computing-enabled networks}. Our proposal is defined by the added capability of optical nodes permitting the superposition of transitional lightpaths for computing purposes to achieve greater capacity efficiency. Specifically, we present two illustrative examples highlighting the potential benefits of bringing about in-network optical computing functions which are relied on optical aggregation and optical XOR gate. The new optical computing capabilities armed at optical nodes therefore call for a radical change in formulating networking problems and designing accompanying algorithms, which are collectively referred to as \textit{optical network design and planning 2.0} so that the capital and operational efficiency could be fully unlocked. To this end, we perform a case study for network coding-enabled optical networks, demonstrating the efficacy of \textit{optical-computing-enabled networks} and the challenges associated with greater complexities in network design problems compared to optical-bypass counterpart.}

\section{Introduction}
\label{intro}
From an architectural perspective, optical networking has been shifted from optical-electrical-optical mode to optical-bypass operations so that transiting lightpaths could be optically cross-connected from one end to the other end rather than being undergone unnecessary optical-electrical-optical conversions \cite{efficient}. Optical-bypass networking has then gone a long way from a conceptual proposal to a widely adopted technology by network operators in the last two decades \cite{all-optical}. However, one of the major challenges in scaling up the optical networks to support explosive traffic growth in a greater efficiency manner is the fact that signal processing functions are mostly implemented in electrical domain based on digital signal processing. This involves a number of the well-established procedures including mainly optical-to-electrical conversion, electronic sampling, digital signal processing and finally, back-conversion to optical domain. Here, the critical bottleneck lies in the electronic sampling rate and in circumventing this major concern, solutions are directed towards migrating certain signal processing functions to the optical domain. Indeed, photonic signal processing / computing technologies offer a new powerful way for handling high speed signals thanks to its inherent merits of wide bandwidth, transparency and energy-efficiency \cite{optical_processing_1}. \\

In optical-bypass networking, the basic functions of optical nodes are to add (drop) and cross-connect the transiting lightpaths. Moreover, in cross-connecting transiting lightpaths through an intermediate node, these lightpaths must be maximally separated from each other in either time, frequency or spatial domain \cite{all-optical} to avoid unwanted interference which deems to deteriorate the signal qualities. This turns out to be a fundamental limitation as various optical computing operations could be performed between such transitional lightpaths to generate the output signals which are spectrally more-efficient than its inputs. In light of recent enormous advances in photonic computing technologies enabling the controlled interference of optical channels for various computing capabilities \cite{agg1, agg2, xor3, nc_others10}, we propose a new architectural paradigm for future optical networks, namely, optical-computing-enabled networks. Our proposal is defined by the added capability of optical nodes permitting the superposition of transitional lightpaths for computing purposes to realize a greater capacity efficiency. Specifically, we present two illustrative examples highlighting the potential benefits of bringing about in-network optical computing which are relied on optical aggregation \cite{agg1, agg2} and optical XOR gate \cite{xor3, nc_others10}. The new optical computing capabilities armed at optical nodes calls for a radical change in optical network design and planning in order to fully reap spectral and cost benefits as well as operational efficiency. To this end, we perform a case study for network coding-enabled optical networks, demonstrating the efficacy of optical-computing-enabled networks and challenges associated with greater complexities in network design problems compared to optical-bypass counterpart.     \\   

The paper is structured as followed. In Section 2, we introduce a new concept of optical-computing-enabled paradigm. We also highlight the applications of two optical computing operations, namely, optical aggregation and optical XOR whose enabling technologies have been progressing fast and their integration to future optical networks could be foreseen. We also address the computational impact and intricacies for network design and planning in the paradigm of optical-computing networking. Next, as a case study to reveal more complicated network design problems arisen in the optical-computing-enabled network, we focus on the network coding-enabled scenarios and formulate the routing, wavelength and network coding assignment problem in Section 3. Section 4 is dedicated to showcase the numerical evaluations comparing our proposal that leverages the use of optical XOR encoding within the framework of optical-computing-enabled networks to the traditional optical-bypass networking. The comparison is drawn on a realistic COST239 and NSFNET network topologies. Finally, Section 5 concludes the paper.

\section{Optical-computing-enabled Paradigm}
The optical-computing-enabled paradigm is characterized by the key property that optical nodes are empowered with the optical computing capability. Specifically, two or more optical channels could be optically mixed together to compute a new optical channel and thus, to attempt achieving a greater capacity efficiency \cite{hai_tnsm, hai_ro, hai_springer3, hai_mttw21}. In light of tremendous progresses in optical computing technologies permitting precisely controlled interference between optical channels, in-transit lightpaths traversing the same optical node are offered unique opportunities to optically superimpose to each other to generate the output signals which are spectrally more-efficient than their inputs. Such optical operations involving the interaction of transitional optical channels pave the way for redefining the optical network architecture, disrupting the conventional assumption of keeping transitional lightpaths untouched. In this context, optical-computing-enabled framework is foreseen to be the next evolution of optical-bypass networking. In this section, we highlight the efficient use and impact of introducing two optical computing operations, namely, optical aggregation (de-aggregation) and optical XOR gate to optical networks. It is noticed that the enabling technologies for realizing such two optical computing capabilities have been accelerating and therefore technical readiness of integration to optical nodes could be foreseen. Of course, there are many other ways for mixing two optical channels and in the future, as photonic computing technologies move forward, a wide range of computing functions could be technologically realized. These advances will be expected to have massive impacts to optical networks from both design, planning, operation and management.  

\subsection{Optical Aggregation and De-aggregation}
Aggregation of lower-speed channels into a single higher-speed one has been a key function in the operation of optical networks. The goal of doing so is to achieve a greater capacity efficiency by freeing up the lightpaths of the lower wavelength utilization. Traditionally, this function has been performed in the electronic domain by terminating optical channels, re-assembling, re-modulating and finally back-converting to the optical domain. Clearly, there have been many limitations of doing so and thus, it is not scalable for higher bit-rate operations. In mitigating this major issue, the concept of optical aggregation has recently been proposed, implemented and pushed forward \cite{agg11, agg12}. The main industrial player for this revolutionary effort is INFINERA, whose the goal is to develop a new ecosystem of devices and components, with the capability of transforming the traditional operation of the optical nodes. In term of functionality, an optical aggregator can add two or more optical channels of lower bit-rate and/or lower-order modulation format into a single higher bit-rate and higher-order modulation format one. In this example, we consider the use of an optical aggregator and de-aggregator whose function is to combine two QPSK signals into a single 16-QAM channel and vice-versa. Fig. 1 illustrates the schematic diagram for adding two QPSK channels of lower bit-rate into a single 16-QAM channel of higher bit-rate. In doing so, the spectral efficiency could thus by improved twice.


\begin{figure}[!ht]
	\centering
	\includegraphics[width=0.6\linewidth, height = 4.5cm]{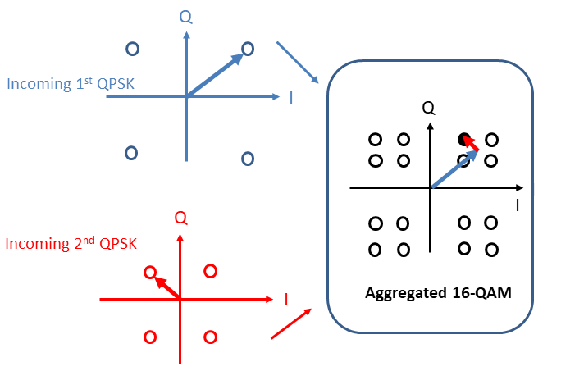}
	\caption{Schematic illustration of the aggregation of two QPSK signals into a single 16-QAM one and vice-versa}
	\label{fig:i1}
\end{figure}

\begin{figure}[!ht]
	\centering
	\includegraphics[width=0.35\linewidth, height = 3cm]{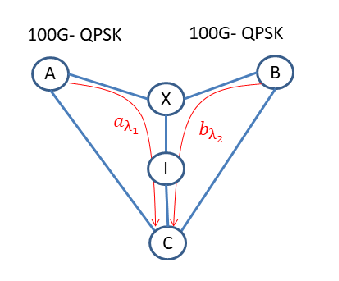}
	\caption{Traffic Provisioning in Optical-bypass Networking}
	\label{fig:i2}
\end{figure}

\begin{figure}[!ht]
	\centering
	\includegraphics[width=0.8\linewidth, height = 4.5cm]{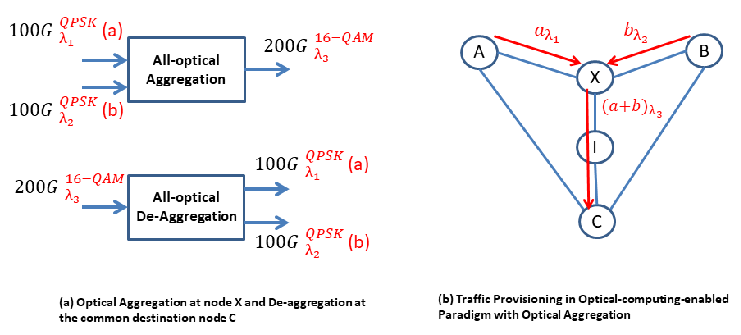}
	\caption{Optical-computing-enabled Paradigm with Optical Aggregation and De-aggregation}
	\label{fig:i3}
\end{figure}

In leveraging the use of the aforementioned aggregator for optical networks, we first consider the conventional way of accommodating traffic demands in optical-bypass networking. Fig. 2 shows the routing and wavelength assignment for two demands $a$ and $b$ of the same line-rate 100G and format QPSK. Due to the wavelength uniqueness constraint on a link, two wavelengths are needed on link $XI$ and $IC$. Now, let us consider the case that at node $X$, the optical aggregation is enabled. By permitting the optical aggregation, two 100G QPSK transitional lightpaths crossing node $X$ could be optically added to generate the output signal of 200G, which is modulated on the 16-QAM format. In Fig. 3, it is clearly observed that by having a single wavelength channel of 200G capacity, a greater capacity efficiency has been realized. At the common destination node $C$, the aggregated lightpath could be decomposed into constituent ones and such decomposition operation could be performed either in an optical or electrical domain. It is important to note that in order to maximize the aggregation opportunities, new network design and planning algorithms should be developed to determine the pairing of demands for aggregation, the respective aggregation node and more importantly, the transmission parameters for aggregated lightpaths. 

\subsection{Optical XOR Encoding and Decoding}
Technologies for realizing all-optical logic gates have been accelerating in recent years that permit performing the bit-wise exclusive-or (XOR) between optical signals of very high bit-rates and/or different modulation formats \cite{xor3, nc_others10}. Different from the aggregation operation, the optical XOR encoding output is kept at the same bit-rate and/or format as the inputs. A functional description of such device is shown in Fig. 5(a), where two optical signals of 100G QPSK on different wavelength are coded together to generate the output X of the same bit-rate and format. 

\begin{figure}[!ht]
	\centering
	\includegraphics[width=0.6\linewidth, height = 3cm]{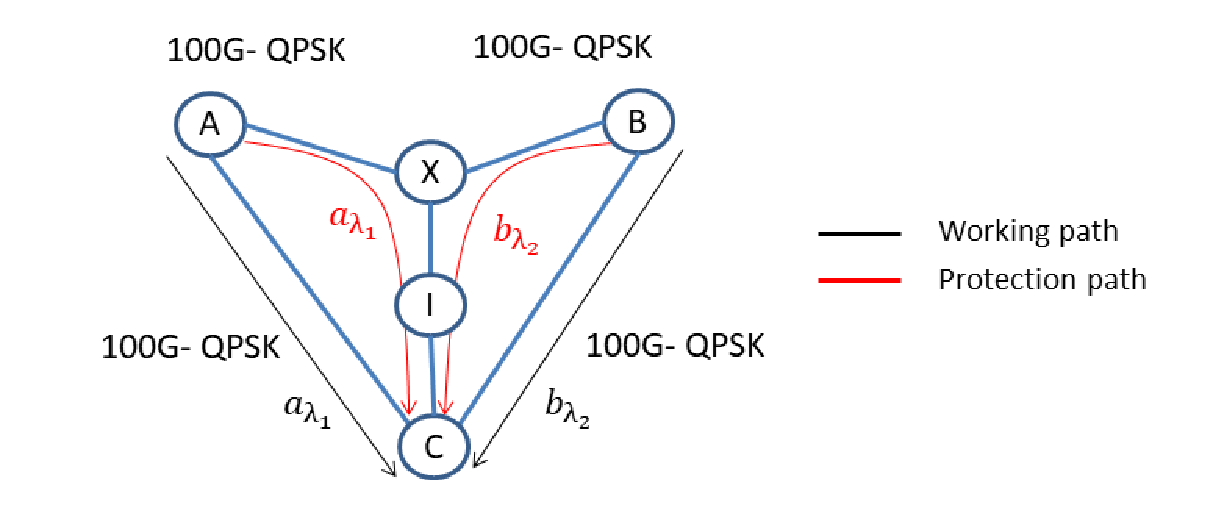}
	\caption{Traffic Provisioning in Optical-bypass Networking}
	\label{fig:i4}
\end{figure}

\begin{figure}[!ht]
	\centering
	\includegraphics[width=0.8\linewidth, height = 4.5cm]{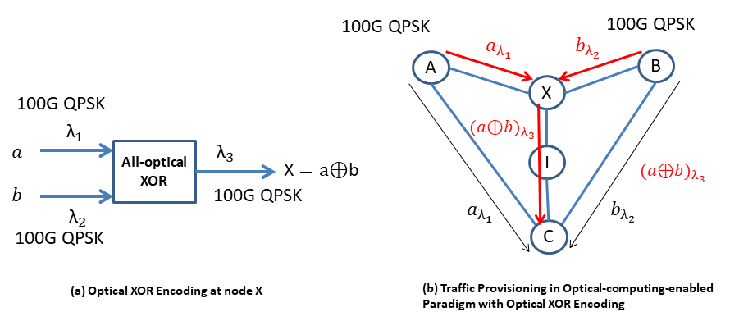}
	\caption{Optical-computing-enabled Paradigm with Optical XOR Encoding and Decoding}
	\label{fig:i5}
\end{figure}

In exploiting the optical XOR gate, we focus on the protection scenario and we assume that there are two demands with dedicated protection. The provisioning of such two demands are shown in Fig. 4 for the optical-bypass framework. Because the protection lightpath of demand $a$ and demand $b$ cross the same links $XI$ and $IC$, it therefore requires at least two wavelengths on those links. Looking at Fig. 5 (b) for the optical-computing-enabled paradigm when node $X$ is armed with the optical XOR encoding capability, the protection signal of demand $a$ and demand $b$ could thus be optically encoded to produce the signal $X = a \oplus b$. Such encoded output is routed all the way from node $X$ to the shared destination node $C$. By doing so, only one wavelength channel on link $XI$ and $IC$ is sufficient, resulting in a spectral saving of $50\%$. Although the protection signals of demand $a$ and demand $b$ is encoded, it is always possible to recover the original signal for both demand $a$ and demand $b$ in any case of a single link failure. The recovery is as simple as the encoding by making use of the XOR operation on the two remaining signals. Specifically, if the working signal of demand $b$ is lost, it can be retrieved in an another way by $b = (a \oplus b) \oplus a$.   \\

The combination of optical encoding and dedicated protection appears to be matched to attain a greater capacity efficiency while keeping a near-immediate recovery speed. Nevertheless, in order to realize the encoding benefits, more complicated network design problems emerge. The more intricacies are related to the determination of the pair of demands for encoding and the selection of the routes and/or transmission parameters for the encoded lightpaths.

\subsection{Impact for Network Design and Planning}
It should be noted that the \textit{optical-computing-enabled} paradigm introduces more networking flexibility by permitting the precisely controlled interference among two or more favorable lightpaths, and this poses important ramifications for the network design algorithms to maximize the potential benefits \cite{hai_comletter, hai_systems, hai_comcom, hai_comcom2, hai_rtuwo, hai_springer2}. In optical-bypass networking, the central problem for the network design and planning is the routing and resource allocation. For solving such problem, the selection of the route and assigning transmission parameters, including the wavelength/spectrum and/or format, is determined for each individual demand. In optical-computing-enabled networking, more complicated network design problems arise due to the interaction of transitional lightpaths. Specifically, in addition to determining of the route and transmission parameters for each demand, the pairing of demands and subsequently, the selection of route and assigning transmission parameters for special lightpaths involving the interaction of two or many demands, must also be identified. This represents a radical change in the network design and planning, disrupting the conventional set of algorithms that have been developed for optical-bypass networking in many years. In recognizing this disruption, we call for a new framework, that is, optical network design and planning 2.0, which encompasses new problems emerging from various ways that transitional lightpaths could be optically mixed and accompanying algorithms including exact and heuristic solutions for solving them. In the subsequent section, we formulate the problem entitled, the routing, wavelength and network coding assignment problem arising in the application of optical XOR for optical-computing-enabled network and highlight how such problem is different from its counterpart, that is, routing and wavelength assignment in optical-bypass network. 

\section{Routing, Wavelength and Network Coding Assignment Problem Formulation in Optical-computing-enabled Network}

In this section, we consider the design of network coding-enabled networks to support a set of traffic demands with minimum wavelength link cost. The optical encoding scheme to be used is the simple XOR, where the input signals of the same wavelength, line-rate and format, are XOR-coded to produce the output signal of the same wavelength and format. The main advantage of such scheme, XOR coding between signals of the same wavelength, is the elimination of a probe signal and therefore, could be highly cost-efficient \cite{xor-model}. Moreover, for ease of operations, the encoding is restricted only on the protection signals of demands having the same destination node and the decoding is only taken place at the destination. Furthermore, each demand is permitted to have maximum one encoding operation. In this framework, there are a set of sufficient constraints on the network coding assignment for any two code-able demands, namely: i) two demands must have common destination,  ii) two demands must use the same wavelength, iii) the link-disjoint constraint between their working paths and between one’s working path to the another’s protection path, iv) their protection paths must have a common sub-path whose one end is the shared destination.  \\

\noindent{Inputs:}
\begin{footnotesize}
	\begin{itemize}
		\item $G(V,E)$: A graph representing the physical network topology with $|V|$ nodes and $|E|$ fiber links. 
		\item $D$: A set representing the traffic demands, indexed by $d$. Each demand $d \in D$ requests \textit{one wavelength capacity} (e.g., 100 Gbps)
		\item $W$: A set representing the wavelengths on each fiber link, indexed by $w$. The link capacity measured in number of wavelength is $|W|$ 
	\end{itemize}
\end{footnotesize}
\noindent{Outputs:}
\begin{footnotesize}
\begin{itemize}
	\item Routing and wavelength assignment for each lightpath
	\item Determination of pair of demands for optical XOR encoding and determination of respective coding nodes.
	\item Routing and wavelength assignment for encoded lightpaths
	\item The usage of wavelength on each link 
\end{itemize}
\end{footnotesize}
\noindent{Objective: \footnotesize{Minimize the wavelength link usage \\}}

The mathematical model is formulated in the form of integer linear programming (ILP). In addition to typical variables and constraints accounting for the selection of route and assigning wavelength for each demand, new variables and constraints emerge as the interaction of demands have been introduced and thus, causing the mathematical model one order of magnitude computationally harder than its counterpart, that is, the traditional routing and wavelength assignment in optical-bypass networking \cite{Algorithm1}. In acknowledging the NP-nature of the model, we therefore propose the following scalable heuristic (Algorithm 1) that could be used in large networks.  

\begin{algorithm}
	\DontPrintSemicolon
	
	\KwInput{$G(V, E), D, W$}
	\KwOutput{$\alpha_{e, w}^{d}$, $\beta_{e, w}^{d}$, $\theta_{w}^{d}$, $z_{e, w}^{d, v}$, $\delta_{v}^{d}$, $f_{d_1}^{d_2}$, $\gamma_{e,w}$} 
	
	\For{node $v \in V$}    
	{ 
		Find demand $d \in D: r(d) = v$ \\
		Insert $d$ into set $X_v$
	}
	
	\textbf{Sort} $X_v$ according to its size $|X_v|$ in descending order  \\
	
	\For{demand $d \in X_v$}    
	{ 
		Find $k$ shortest cycles (modified Suurballe algorithm) including working and routing route for demand $d$ \\
	}
	
	\textbf{Sort} demand $d \in X_v$ according to its length of $k$ cycles in descending order \\
	
	\tcc {Perform routing, encoding and wavelength assignment for all sorted demands}
	
	\For {$ d \in X_v$}  
	{ Select one cycle for demand $d$ out of $k$ cycles \\
		Perform encoding with a suitable demand \\
		First-fit Wavelength Assignment \\
	}
	
	%
	%
	\caption{Heuristic Solution}
\end{algorithm}

\section{Numerical Results}
\begin{figure}[!ht]
	\centering
	\includegraphics[width=0.7\linewidth, height = 6cm]{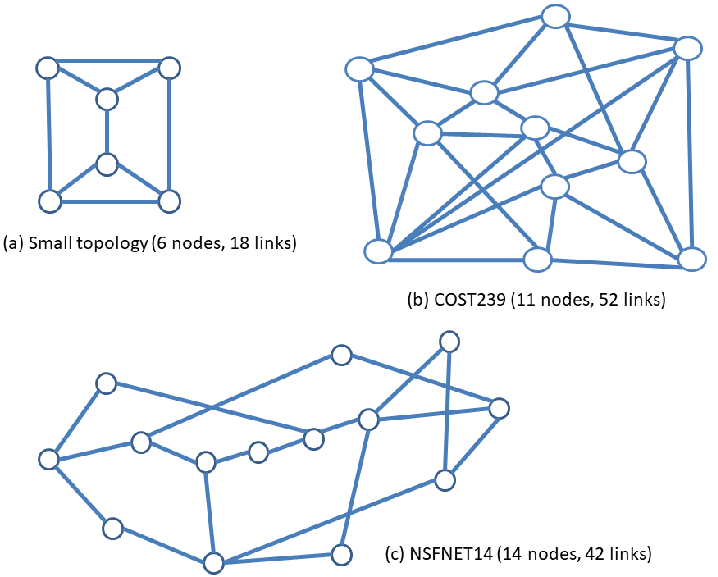}
	\caption{Network Topologies under Test}
	\label{fig:topo}
\end{figure}

This section presents the numerical evaluations comparing our proposal that leverages the use of optical XOR encoding within the framework of optical-computing-enabled networks to the traditional optical-bypass networking. The comparison is drawn on a realistic COST239 and NSFNET network topologies as shown in Fig. \ref{fig:topo}. The metric for comparison is a traditional one, that is, the wavelength link cost. Two designs, namely, WNC and NC are performed, where WNC refers to the routing and wavelength assignment in optical-bypass networking and NC refers to the more advanced problem, i.e., routing, wavelength and network coding assignment in optical-computing-enabled networks.  \\

We first test the solutions from solving the ILP models and heuristic on small-scale topology, 6 nodes, all of degree 3 as shown in Fig. 6 (a). The traffic is randomly generated between node pairs with the unit capacity (i.e., one wavelength) and the fiber capacity is 40 wavelengths. We consider three increasing load corresponding to $30\%$, $70\%$ and $100\%$ (full mesh) node-pair traffic exchange. The result in Table 1 (except the full mesh) is averaged over 20 samples. For the network coding-based design in full mesh traffic, the computation is overly long and thus, the results is obtained after 10 hours of running. The well-studied heuristic for WNC achieves optimal results which are on a par with its ILP model while the heuristic for NC produces reasonably good solutions with tight gaps compared to its ILP model, avoiding the overly long computational time. Due to the sub-optimal nature of the heuristic algorithms, the gain obtained by these algorithms is slightly reduced compared to the one obtained from the ILP. \\

\begin{table*}[!ht]
	\caption{Performance Comparison between Exact Solution and Heuristic One}
	\label{tab: r1}
	\centering
	\begin{tabular}{|c|c|c|c|c|c|c|}
		\hline
		\multirow{2}{*}{Load} & \multicolumn{3}{|c|}{ILP}  & \multicolumn{3}{|c|}{Heuristic} \\
		\cline{2-7}
		& WNC & NC & Gain & WNC & NC & Gain\\ 
		\hline
		$30 \%$ & $32.6$ & $30.6$ & Max: $9\%$, Mean: $6\%$ & $32.6$  & $31.2$ & Max: $9\%$, Mean: $4\%$\\
		\hline
		$70 \%$ & $75.8$ & $67.9$ & Max: $12\%$, Mean: $10\%$ & $75.8$ & $70.6$ & Max: $9\%$, Mean: $7\%$\\
		\hline
		$100\%$ & $108$ & \underline{$99$} & Max=Mean= $8\%$ & 108 & 100 & Max=Mean= $7\%$\\
		\hline
	\end{tabular}
\end{table*}

We apply the heuristic algorithm for larger networks, NSFNET and COST239 topologies with the same setting as in the 6-node topology about traffic generation, fiber capacity and number of traffic samples. The results are presented in Table 2. It can be observed that up to about $8\%$ gain could be achieved with the NSFNET network. For more densely connected COST239 network, the lower gain is obtained, up to $5\%$. It is evident that the solution from NC cases is always better than that from the WNC, resulting in improved capacity efficiency. Compared to the findings on O-E-O case \cite{icc}, where the gain was reported up to $20\%$, there is reduced gain in all-optical case. This may be due to wavelength-related constraints for network coding assignments, curbing the coding capability among the demands. Moreover, it should be noted that the gain is highly dependent on the structure of the network topology, traffic and network design algorithms. 

\begin{table*}[!ht]
	\centering
	\caption{Numerical Results for Realistic topologies}
	\begin{tabular}{|c|c|c|c|c|c|}
		\hline
		Topo & Load & WNC & NC & Gain & No Coding Operation \\
		\hline
		\multirow{3}{*}{NSFNET} & $30\%$ & 318.3 & 299.5 & Max = $8\%$, Mean = $6\%$ & 9.7 \\
		\cline{2-6}
		& $70\%$ & 730.4 & 682.4 & Max = $8\%$, Mean = $7\%$ & 24.5 \\
		\cline{2-6}
		& $100\%$ & 1048 & 981 & Max = Mean = $6\%$ & 35 \\
		\hline
		\multirow{3}{*}{COST239} & $30\%$ & 126.2 & 123.3 & Max = $3\%$, Mean = $2\%$ & 1.4 \\
		\cline{2-6}
		& $70\%$ & 295.6 & 285.4 & Max = $5\%$, Mean = $3\%$ & 5.1\\
		\cline{2-6}
		& $100\%$ & 420 & 404 & Max = Mean = $4\%$ & 8 \\
		\hline
	\end{tabular}
	
	\label{table:ta}
\end{table*}

\section{Conclusion}
This paper presents a new networking paradigm for future optical networks named, \textit{optical-computing-enabled framework}. As a potential candidate for the next evolution of optical-bypass architecture, our proposal aims to exploit the optical computing capability at optical nodes so that greater capacity efficiency could be achieved. In highlighting the potential benefits of such \textit{optical-computing-enabled framework}, we brought out two revealing examples leveraging the efficient use of optical aggregation and optical XOR gate. A numerical case study for the network coding-enabled design was provided to demonstrate the efficacy and the more complicated network design algorithm of \textit{optical-computing-enabled networking} compared to the optical-bypass counterpart. \\

Albeit still primitive, the perspective of permitting optical mixing at intermediate nodes heralds a reinvention of optical networking, altering the way we think about optical network architecture. Although the experiments and practical demonstration of optical-computing-enabled networks are still ahead, the disconnection between the theoretical studies and the implementation realities are only just beginning to be rectified as enabling technologies have been more maturing and the huge profits have been foreseen. It should be pointed out that network design algorithms play a key role in achieving the operational efficiency of a network and thus, more robust and carefully designed algorithms should be developed to optimize the advantages of \textit{optical-computing-enabled networking}. As for future works, we plan to develop an ecosystem of new research problems and the accompanying algorithms, collectively referred to as \textit{optical network design and planning 2.0}, to capture a wide range of optical computing operations between in-transit lightpaths at their optimal usage scenarios. 


\bibliographystyle{spmpsci_modified}      
\bibliography{ref}   

\end{document}